\documentclass[12pt, preprint]{aastex}

\usepackage{amsmath}
\usepackage{graphicx}
\usepackage{color}

\newcommand{\be}{\begin{eqnarray}}
\newcommand{\ee}{\end{eqnarray}}

\shorttitle{Testing black hole candidates with iron lines}
\shortauthors{Jiang et al.}

\begin{document}

\title{Testing the Kerr Nature of Black Hole Candidates using Iron Line Spectra in the CPR Framework}

\author{Jiachen Jiang\altaffilmark{1}, Cosimo Bambi\altaffilmark{1}, and James F.\ Steiner\altaffilmark{2}\altaffilmark{\dag}\\}

\altaffiltext{1}{Center for Field Theory and Particle Physics and Department of Physics, 
Fudan University, 200433 Shanghai, China}
\altaffiltext{2}{Harvard-Smithsonian Center for Astrophysics, Cambridge, MA 02138, United States}
\altaffiltext{\dag}{Hubble Fellow}
\email{Corresponding author: bambi@fudan.edu.cn}

\begin{abstract}
The iron K$\alpha$ line commonly observed in the X-ray spectrum of both stellar-mass and supermassive black hole candidates originates from X-ray fluorescence of the inner accretion disk.  Accordingly, it can be used to map the spacetime geometry around these objects. In this paper, we extend previous work using the iron K$\alpha$ line to test the Kerr black hole hypothesis. We adopt the Cardoso-Pani-Rico parametrization and we test the possibility of constraining possible deviations from the Kerr solution that can be obtained from observations across the range of black hole spins and inclination angles. We confirm previous claims that the iron K$\alpha$ line is potentially a quite powerful probe for testing the Kerr metric given sufficiently high quality data and with systematics under control, especially in the case of fast-rotating black holes and high inclination angles since both conditions serve to maximize  relativistic effects. We find that some geometric perturbations from Kerr geometry manifest more strongly in the iron line profile than others.  While the perturbation parameter $\epsilon^t_3$ can be well constrained by the iron line profile, an orthogonal data set is necessary to constrain departures from Kerr geometry in $\epsilon^r_3$.  

\end{abstract}



\section{Introduction \label{s-1}}

When a star exhausts its nuclear fuel, the core shrinks to find a new equilibrium configuration. In the case of a sufficiently massive star, there is no known mechanism capable of providing the necessary pressure support to halt the process, and the body undergoes complete collapse. In general relativity and under a set of reasonable assumptions, the final product of the collapse is a black hole (BH)~\citep{luca1,luca2,giacomazzo,collapse}. In 4 dimensions, the only uncharged BH solution of the vacuum Einstein's equations is the Kerr metric. The spacetime around astrophysical BHs should be well described by the Kerr geometry, since the presence of the accretion disk is usually completely negligible, given that the disk mass is many orders of magnitude smaller than that of the BH~\citep{disk}.

The objects termed BH candidates are dark and compact and are naturally interpreted as the Kerr BHs of general relativity. Many of these sources have robust mass measurements, obtained dynamically (e.g., \citealt{o10}, and references therein).  When a compact object in an X-ray binary is more massive than $\approx 3~M_\odot$, the object is classified as a BH candidate, since it exceeds the maximum mass for a neutron star~\citep{rr}. With a similar spirit, the supermassive bodies at the center of every normal galaxy are BH candidates because they are too massive, compact, and too old to be clusters of neutron stars\footnote{The cluster lifetime due to evaporation and collisions is less than the age of these systems \citep{maoz}.}. The absence of thermal radiation emitted from the surface of BH candidates may also be interpreted as evidence for the presence of a horizon (\citealt{h1_a, h1_b}; but see \citealt{h2_a,h2_b}). Eventually, the Kerr paradigm relies on the validity of general relativity in strong gravitational fields. However, deviations from the Kerr solution may be expected from a number of theoretical arguments, see, e.g.~\citet{motiv_a, motiv_b,motiv_c}. The experimental confirmation of the Kerr nature of BH candidates is thus an important issue in the fields of astrophysics, gravity, and high energy physics~\citep{review_a, review_b}.

The spacetime geometry around BH candidates can be potentially tested by studying the properties of the electromagnetic radiation emitted by the gas in the accretion disk. Such radiation is inevitably affected by relativistic effects and therefore carries information about the metric. Accurate measurements of the thermal spectrum of a thin disk~\citep{cfm-k, mns} and of the iron K$\alpha$ line~\citep{iron-k, iron-r} are today the principal techniques for providing information about the nature of BH candidates and they can be used to test the Kerr metric~\citep{cb1_a, cb1_b, cb1_c,cb2_a, cb2_b,cb2_c,cb2_d,jp,torres_a,torres_b}. However, the usual problem with such a test is often strong degeneracy among system parameters, in particular between the parameter measuring deviation from Kerr geometry, and the spin and the inclination angle. The thermal spectrum of thin disks has a simple shape and therefore it is fundamentally more challenging to break such a degeneracy~\citep{cfm-nk_a, cfm-nk_b}. The iron line profile is potentially more powerful. With available X-ray data, it has already been established that BH candidates cannot be some types of boson stars and of traversable wormholes~\citep{iron-nk_a, iron-nk_b}. However, it isn't clear whether present data are sufficiently precise to test the Kerr metric; instead, this is a target for the next generation of X-ray satellites~\citep{jjc}.


The aim of this paper is to extend previous studies which test the Kerr metric using of the iron K$\alpha$ line. In what follows, we employ the simplest version of the Cardoso-Pani-Rico parametrization and we consider two types of deformation in their schema, $\epsilon_3^t$ and $\epsilon_3^r$~\citep{cpr}. We discuss constraining $\epsilon_3^t$ and $\epsilon_3^r$ using iron lines for different values of the BH spin and for a range of disk inclination angles with respect to our line of sight. We find that the best source with which to test the Kerr metric has high spin and a large inclination. Indeed, for a fast-rotating Kerr BH, the inner edge of the disk extends very close to the compact object, increasing the iron line's ability to probe the region in which deviations from the Kerr geometry are most likely to be pronounced. A large inclination angle can likewise maximize the light bending effects, and accordingly amplify any differences resulting from alternate spacetimes. Some modes of deformation are surely more readily constrained than others. The deformation parameter $\epsilon_3^t$ principally alters the position of the inner edge of the disk, whose effects on the iron line profile is generally more prominent than a non-vanishing $\epsilon_3^r$, whose main effect is to upon photon trajectories.  The best candidates source to test for these variations would be a system like GRS1915+105, which is quite bright and harbors what appears to be a fast-rotating Kerr BH at a relatively large inclination angle ($a_* > 0.98$ and $i = 60^\circ$; ~\citealt{grs1915, reid}).

Throughout the Paper, we employ units in which $G_{\rm N} = c = 1$ and the convention of a metric with signature $(-+++)$.

\section{Testing the Kerr paradigm \label{s-2}}

Gas in the accretion disk and the photons emitted by the disk can be treated as test-particles and are assumed to follow the spacetime geodesics; this is always the case in a metric theory of gravity but is not so in more general frameworks. The adopted approach is accordingly only sensitive to the geodesics of the metric and it cannot completely test the Einstein equations, in the sense that it is fundamentally impossible to use such a technique to distinguish a Kerr BH of general relativity from a Kerr BH in an alternative metric theory of gravity~\citep{knk}. Bearing this in mind, the usual strategy is to employ an approach similar to the Parametrized Post-Newtonian (PPN) formalism~\citep{will}, which has been used to test the Schwarzschild solution in the weak field limit within the Solar System since the 1960s. In the PPN case, one writes the most general static and spherically symmetric solution in a weak gravitational field, by using an expansion in $M/r$. In isotropic coordinates, the line element turns out to be
\be\label{eq-ppn}
ds^2 = - \left(1 - \frac{2 M}{r} + \beta\frac{2 M^2}{r^2} + . . .  \right) dt^2 
+ \left(1 + \gamma \frac{2 M}{r} + . . .  \right) \left(dx^2 + dy^2 + dz^2 \right) \, ,
\ee
where $\beta$ and $\gamma$ parametrize our ignorance, while the term $2M/r$ in $g_{tt}$ is necessary to have the correct Newtonian limit\footnote{The PPN approach is traditionally formulated in isotropic coordinates and the line element of the most general static and spherically symmetric distribution of matter is given by Eq.~(\ref{eq-ppn}). When cast in the more familiar Schwarzschild coordinates, the line element becomes
\be
ds^2 = - \left(1 - \frac{2 M}{r} + \left(\beta - \gamma\right)\frac{2 M^2}{r^2} + . . .  \right) dt^2 
+ \left(1 + \gamma \frac{2 M}{r} + . . .  \right) dr^2 + r^2 d\theta^2 + r^2 \sin^2\theta d\phi^2 \, .
\ee}. In general relativity, we know that the only static and spherically symmetric vacuum solution is the Schwarzschild metric, and cast in the form above it has $\beta = \gamma = 1$. Since we want to test the Schwarzschild metric, we leave $\beta$ and $\gamma$ free and we aim to constrain the two parameters with observations. Today's best measurements are~\citep{ppn_a, ppn_b}
\be
|\beta - 1| < 2.3 \cdot 10^{-4} \, , \quad
|\gamma - 1| < 2.3 \cdot 10^{-5} \, .
\ee

To test the Kerr metric, we can use a similar approach, namely we consider a more general solution that includes the Kerr metric as special case. Such a metric will be described by a set of ``deformation parameters'', whose value will be determined by observations and the Kerr geometry is recovered when all the deformation parameters vanish. However, now we want to test the Kerr metric near the BH, so we cannot use an expansion in $M/r$ any longer.  Spacetime is axisymmetric, not spherically symmetric, and if we try to alter by hand the Kerr metric it is difficult to avoid creating a spacetime with pathological features like naked singularities and regions with closed time-like curves. For the time being, there is not yet a standard formalism for testing the Kerr metric, even if some proposals are present in the literature~\citep{jp-metric,cpr,zhidenko}. In what follows, we adopt the Cardoso-Pani-Rico parametrization~\citep{cpr}. In Boyer-Lindquist coordinates, the line element reads
\be\label{eq-m}
ds^2 &=& - \left(1 - \frac{2 M r}{\Sigma}\right)\left(1 + h^t\right) dt^2
+ \frac{\Sigma \left(1 + h^r\right)}{\Delta + h^r a^2 \sin^2\theta} dr^2
+ \Sigma d\theta^2 \nonumber\\ 
&& - 2 a \sin^2\theta \left[\sqrt{\left(1 + h^t\right)\left(1 + h^r\right)} 
- \left(1 - \frac{2 M r}{\Sigma}\right)\left(1 + h^t\right)\right] dt d\phi \nonumber\\
&& + \sin^2\theta \left\{\Sigma + a^2 \sin^2\theta \left[ 2 \sqrt{\left(1 + h^t\right)
\left(1 + h^r\right)} - \left(1 - \frac{2 M r}{\Sigma}\right)
\left(1 + h^t\right)\right]\right\} d\phi^2 \, , \quad
\ee
where $a = J/M$ is the BH spin parameter, $J$ is the BH spin angular momentum, $\Sigma = r^2 + a^2 \cos^2 \theta$, $\Delta = r^2 - 2 M r + a^2$, and
\be
h^t = \sum_{k=0}^{+\infty} \left(\epsilon_{2k}^t 
+ \epsilon_{2k+1}^t \frac{M r}{\Sigma}\right)\left(\frac{M^2}{\Sigma}
\right)^k\, , 
\qquad
h^r = \sum_{k=0}^{+\infty} \left(\epsilon_{2k}^r
+ \epsilon_{2k+1}^r \frac{M r}{\Sigma}\right)\left(\frac{M^2}{\Sigma}
\right)^k \, .
\ee
There are two infinite sets of deformation parameters, $\{\epsilon_k^t\}$ and $\{\epsilon_k^r\}$ (at increasingly high order).  The line element in~(\ref{eq-m}) reduces to the Johannsen-Psaltis one for $h^t = h^r$~\citep{jp-metric}, and to the Kerr line element when $h^t = h^r = 0$. In the Johannsen-Psaltis background, $\epsilon_0 = 0$ gives an asymptotically flat spacetime, while $\epsilon_1$ and $\epsilon_2$ must be small to meet the Solar System constraints~\citep{cpr}. $\epsilon_3$ is the first unbounded deformation parameter and there are no qualitative differences between $\epsilon_3$ and higher order terms, in the sense that all these parameters deform the spacetime geometry in a similar way and leave the same observational signature in the spectrum~\citep{agn}. In what follows, we only consider the possibility of non-vanishing $\epsilon_3^t$ and $\epsilon_3^r$. These deformation parameters introduce deviations from the Kerr geodesic motion, which can be seen as a departure from the predictions of Einstein's theory of gravity on how matter distorts the nearby spacetime.

\begin{figure}
\begin{center}
\vspace{0.5cm}
\includegraphics[type=pdf,ext=.pdf,read=.pdf,width=7.8cm]{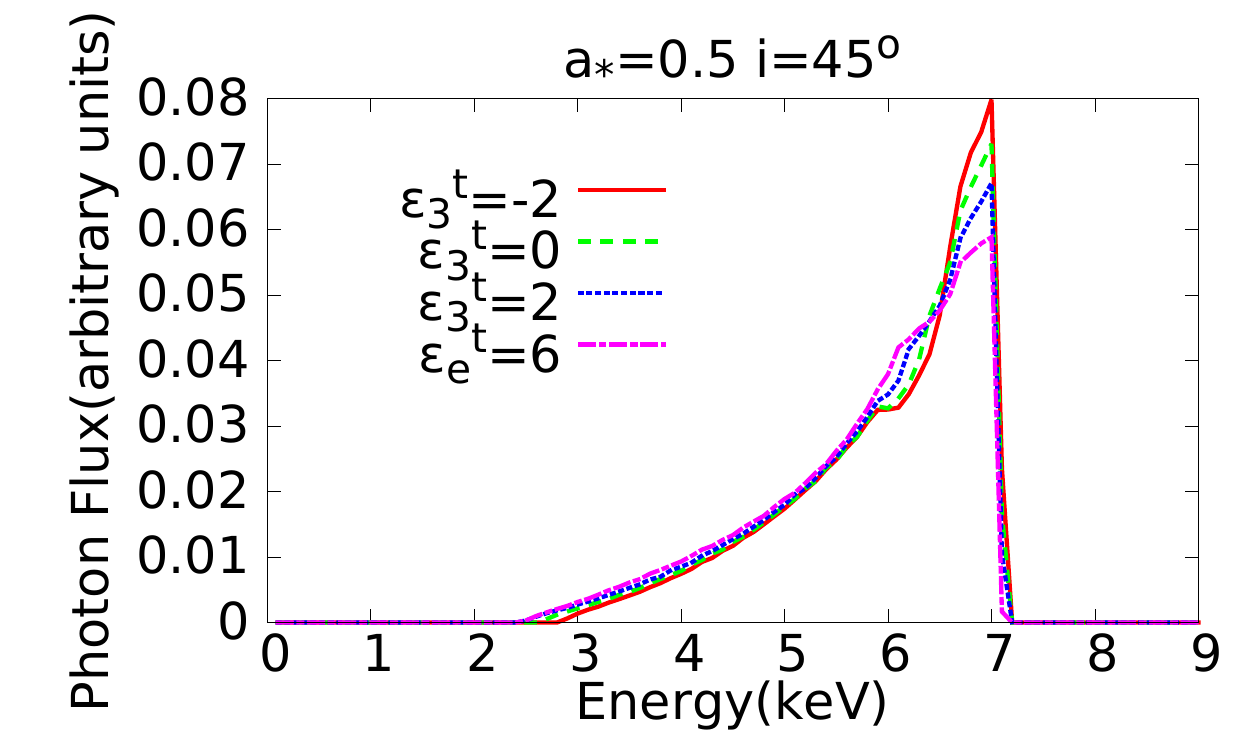}
\hspace{0.5cm}
\includegraphics[type=pdf,ext=.pdf,read=.pdf,width=7.8cm]{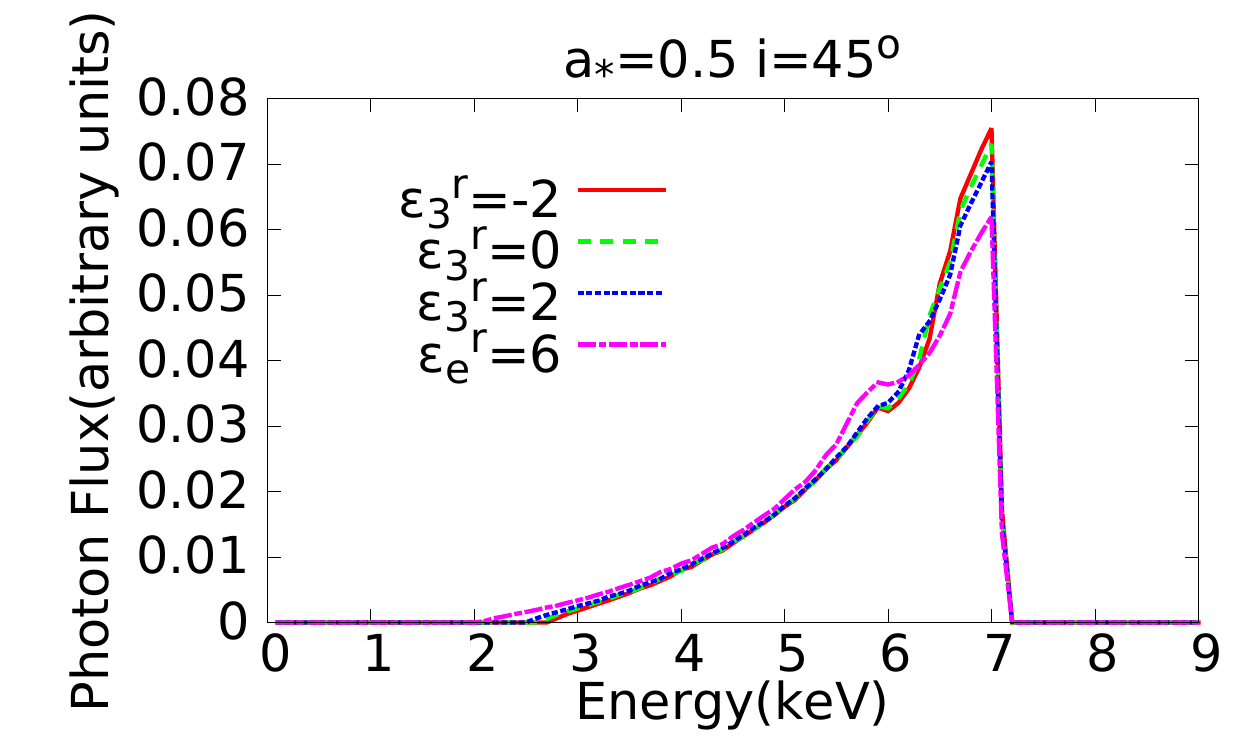} \\
\vspace{0.8cm}
\includegraphics[type=pdf,ext=.pdf,read=.pdf,width=7.8cm]{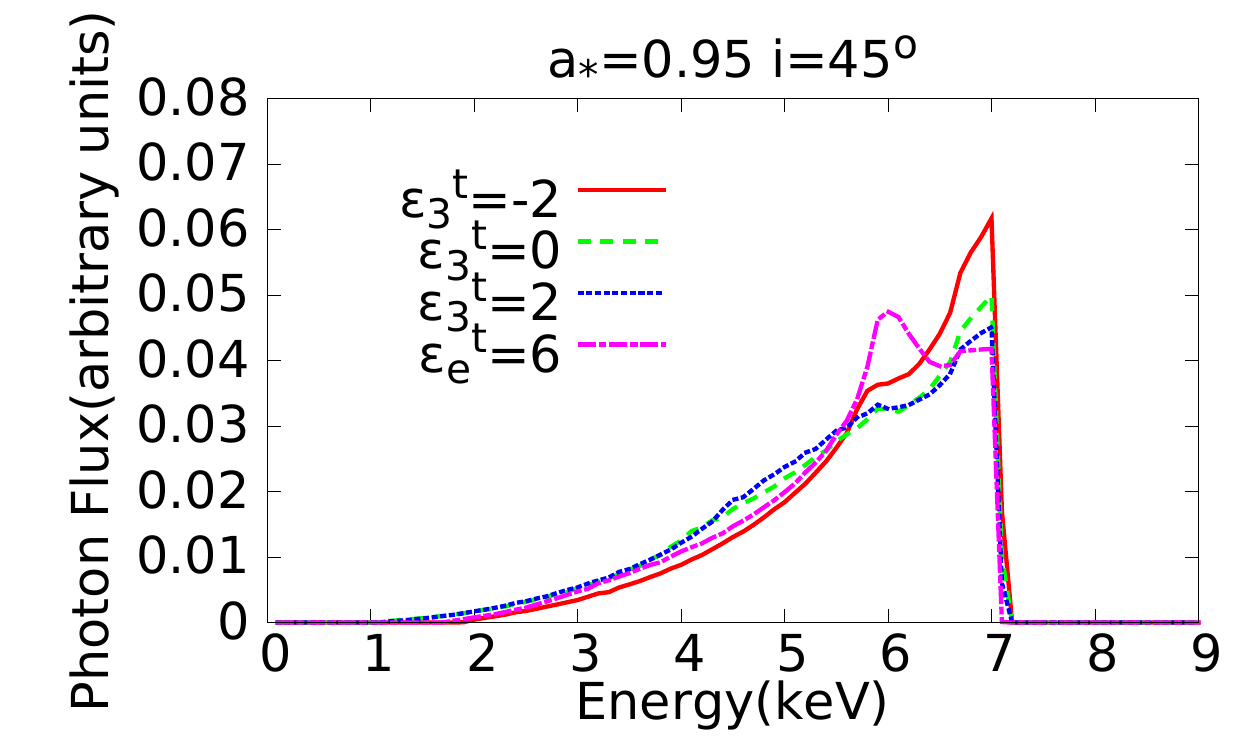}
\hspace{0.5cm}
\includegraphics[type=pdf,ext=.pdf,read=.pdf,width=7.8cm]{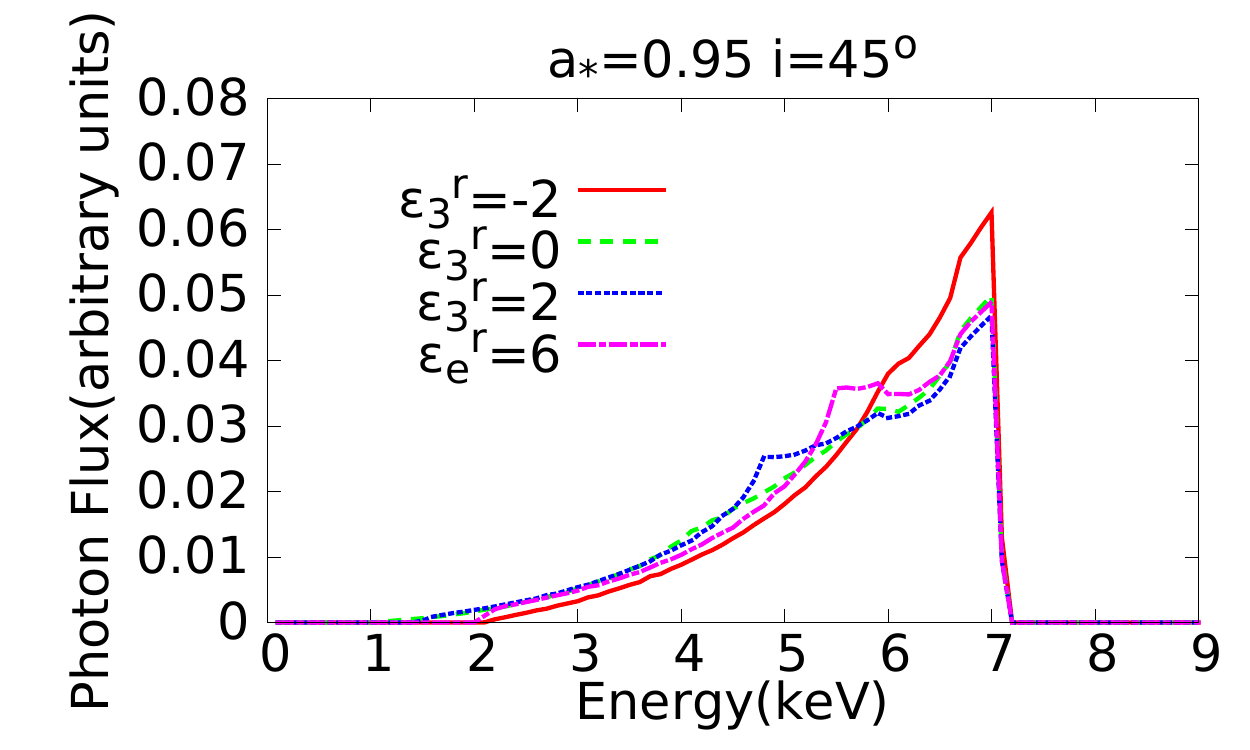}
\end{center}
\caption{Impact of the deformation parameters $\epsilon^t_3$ and $\epsilon^r_3$ on the profile of the iron line. The spin parameter is $a_* = 0.5$ in the top panels and $a_* = 0.95$ in the bottom panels. The viewing angle is always $i=45^\circ$. The left panels show the change which results from varying $\epsilon^t_3$ while keeping $\epsilon^r_3 = 0$. In the right panels, $\epsilon^t_3 = 0$ and we instead vary $\epsilon^r_3$. }
\label{fig0}
\end{figure}

\section{Simulations \label{s-3}}

The broad iron K$\alpha$ line observed in the X-ray spectrum of BH candidates of all masses originates when hard X-rays produced in a corona illuminate the accretion disk, fluorescing the line emission~\citep{iron-r}. Geometrically thin, optically thick accretion disks can be described by the Novikov-Thorne model~\citep{nt-m_a, nt-m_b}, in which the gas particles follow geodesic circular orbits in the equatorial plane and the inner edge of the disk extends inwards to the innermost stable circular orbit, an assumption supported empirically by the stability of the measured inner edge of the disk across time and over an order of magnitude of luminosity variation~(e.g., \citealt{isco}). Calculating the iron line profile for a non-Kerr background is straightforward and discussed elsewhere (e.g., \citealt{cb2_a, cb2_b, cb2_c, cb2_d,jp}). The effect of the deformation parameters $\epsilon^t_3$ and $\epsilon^r_3$ on the iron line profile is shown in Fig.~\ref{fig0}. A non-vanishing deformation parameter mainly affects the extension of the low energy tail of the iron line profile (as it changes the inner edge of the disk) as well as the photon redshift at 5-20 gravitational radii (which affects the shape of the iron line at high energies).

In this section, we compute the iron line assuming a Kerr background while estimating plausible constraints on $\epsilon^t_3$ and $\epsilon^r_3$ to be obtained from observations. In doing so, we simulate the effect of observing our model in alternate backgrounds to assess how readily features can be distinguished at a given level of confidence. In this work, we use the approach employed in~\citet{jjc}. We adopt a primed notation (e.g., $a_*'$, $i'$) for the input model from simulated data are derived.  Unprimed terms are used to refer to the model employed by the spectral fitting routine.

Our simulation consists of first computing the photon flux of the iron line profile over a grid of energy bins with resolution $\Delta E = 100$~eV from 1--9~keV.  For the sake of simplicity, the emissivity index for both the simulation and in the model fitting is always assumed to be 3; namely the intensity profile scales as $1/r^3$. Such a value correspond to the Newtonian limit for lamppost coronal geometry (e.g., \citealt{fabian_2012}).  A more exact model consisting of a radially evolving emissivity index -- $q$ -- may be useful when treating real data (see, e.g., \citealt{dauser}); however, in this exploratory work we adopt a widely-used simplification $q=3$ in both generating our simulated data, and in its analysis. In our simulations, the iron line profile is added to a power-law continuum. The latter is normalized to include 100 times the number of iron line photons when integrated over the energy range 1--9~keV, and is generated using a photon index $\Gamma' = 2$.  Depending on the line shape, this corresponds to a line equivalent width of $EW \approx 370-440$~eV.  In the analysis below, the ratio between the continuum and the iron line photon flux, $K$, as well as the index of the continuum, $\Gamma$, are left as free parameters to be determined by the fit.

We employ the standard analysis technique common throughout X-ray astronomy: our simulated spectra include Poisson noise, are then binned to achieve a threshold of counts, and fitted employing $\chi^2$ analysis. The Poisson noise is trivial to include; in each bin it is a randomly generated realization using the ``true'' model prediction for the photon count.  The simulated (noisy) photon count number, is what we employ as the data for our analysis.   The level at which the noise manifests depends on the total number of spectral counts; in essence the total counts therefore set the scale of the noise in the spectrum. In this paper, we assume that the number of iron-line photons is $N = 10^4$, corresponding approximately to the {\em line} signal expected in a next generation high-quality observation of a supermassive black hole in an active galactic nucleus (AGN). To treat the simulation as if it were data, our analysis employs standard techniques, consisting of grouping the data to a minimum of 20 counts per bin.  Our results are not sensitive to this particular choice of threshold.

\begin{figure}
\begin{center}
\vspace{0.5cm}
\includegraphics[type=pdf,ext=.pdf,read=.pdf,width=7.0cm]{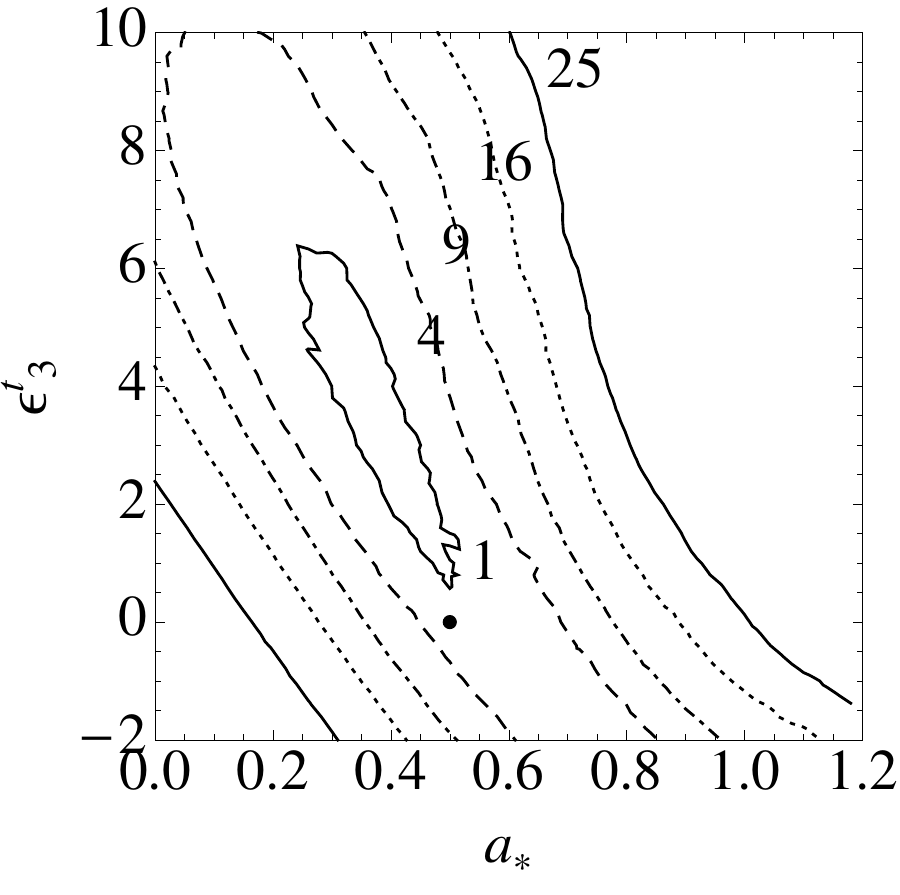}
\hspace{0.5cm}
\includegraphics[type=pdf,ext=.pdf,read=.pdf,width=7.0cm]{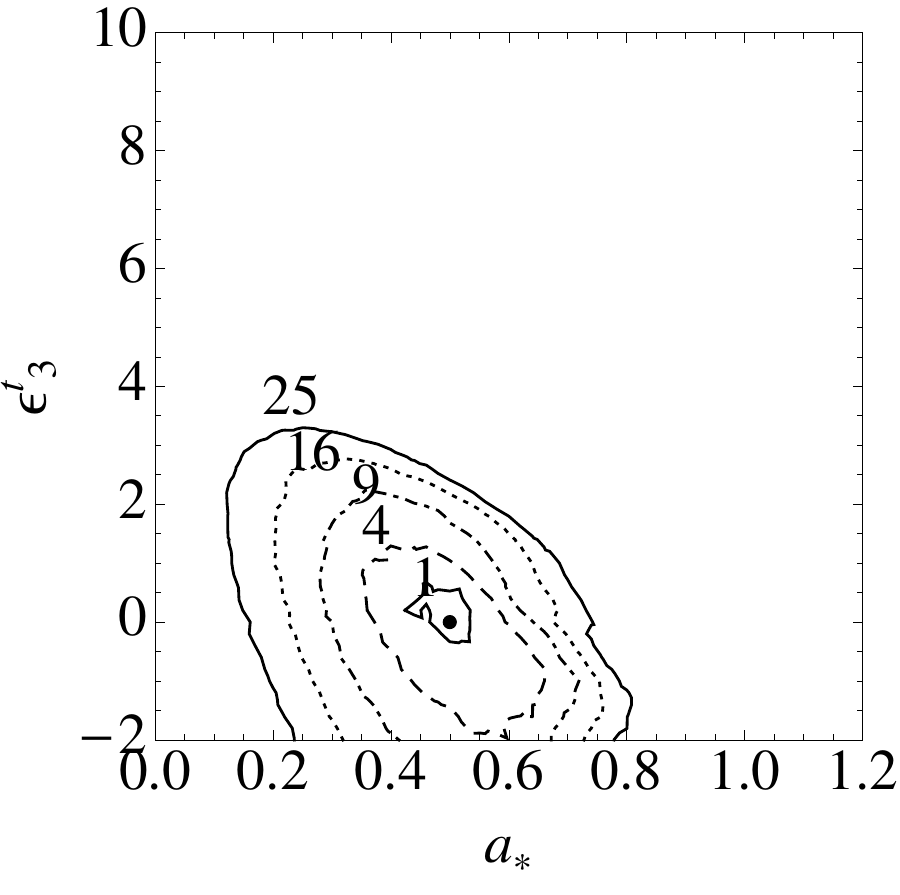} \\
\vspace{0.8cm}
\includegraphics[type=pdf,ext=.pdf,read=.pdf,width=7.0cm]{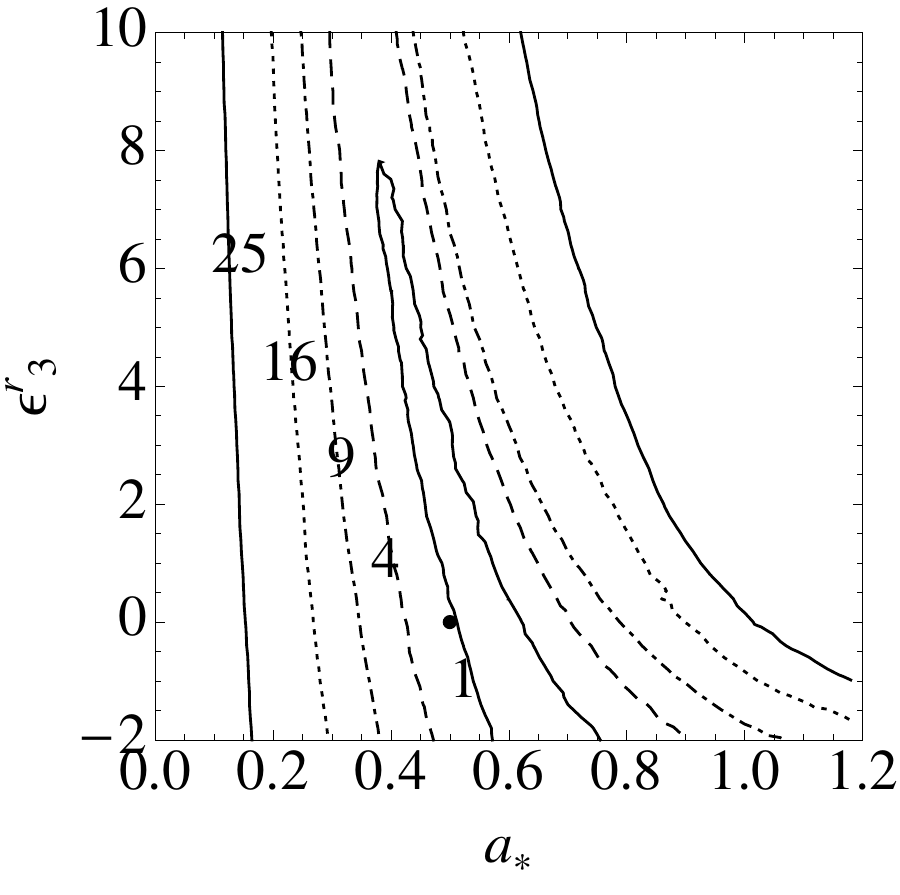}
\hspace{0.5cm}
\includegraphics[type=pdf,ext=.pdf,read=.pdf,width=7.0cm]{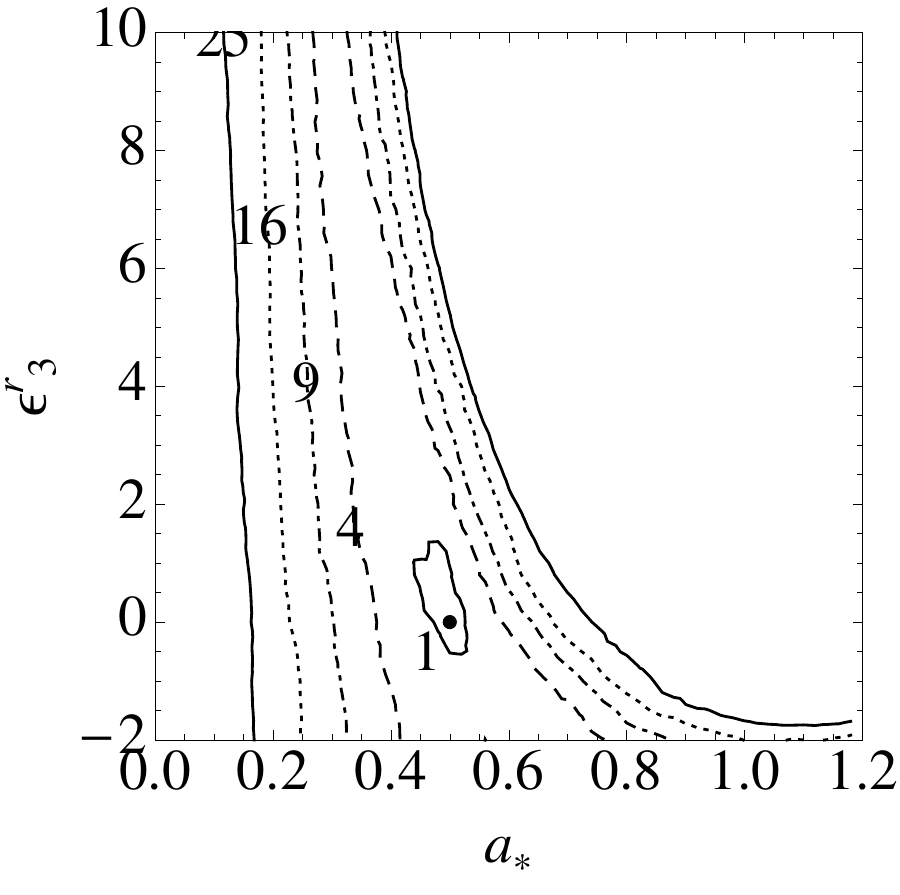}
\end{center}
\caption{Top panels: $\Delta \chi^2$ contours with $N=10^4$ from the comparison of the iron line profile of a Kerr BH simulated using an input spin parameter $a_*' = 0.5$ and an inclination of $i'=20^\circ$ (left panel) and $i' = 70^\circ$ (right panel) versus a set of Cardoso-Pani-Rico BHs with spin parameters $a_*$ and a non-vanishing deformation parameter $\epsilon^t_3$. Bottom panels: as in the top panels for a non-vanishing deformation parameter $\epsilon^r_3$.}
\label{fig1}
\end{figure}

\begin{figure}
\begin{center}
\vspace{0.5cm}
\includegraphics[type=pdf,ext=.pdf,read=.pdf,width=7.0cm]{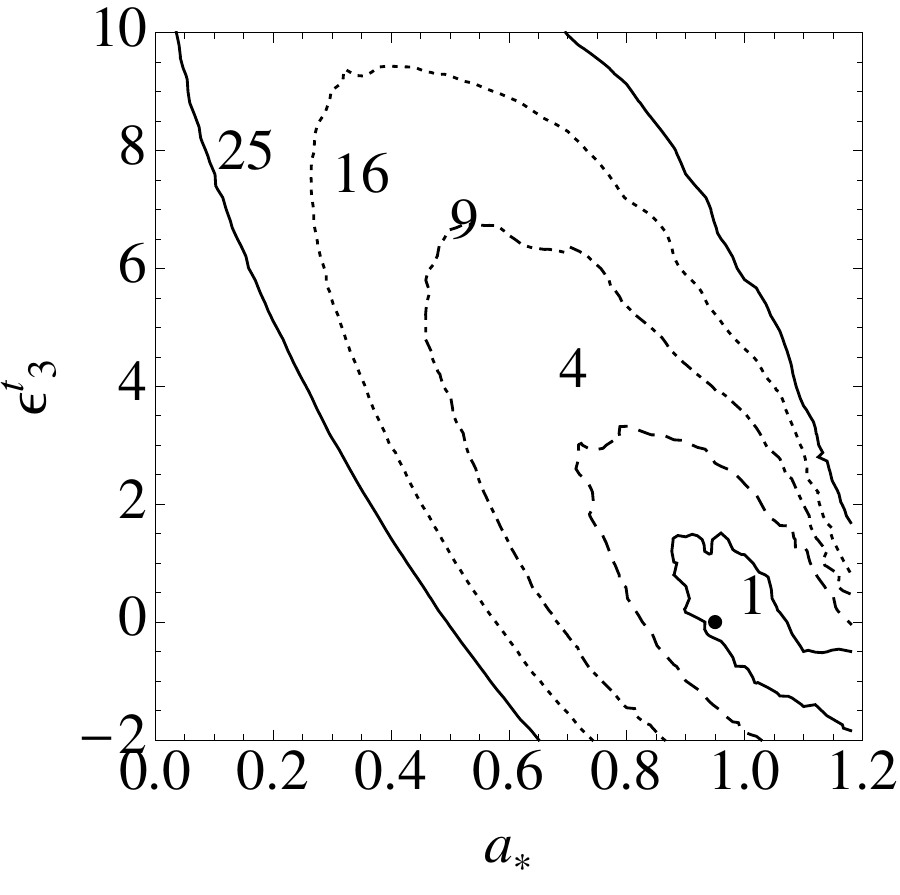}
\hspace{0.5cm}
\includegraphics[type=pdf,ext=.pdf,read=.pdf,width=7.0cm]{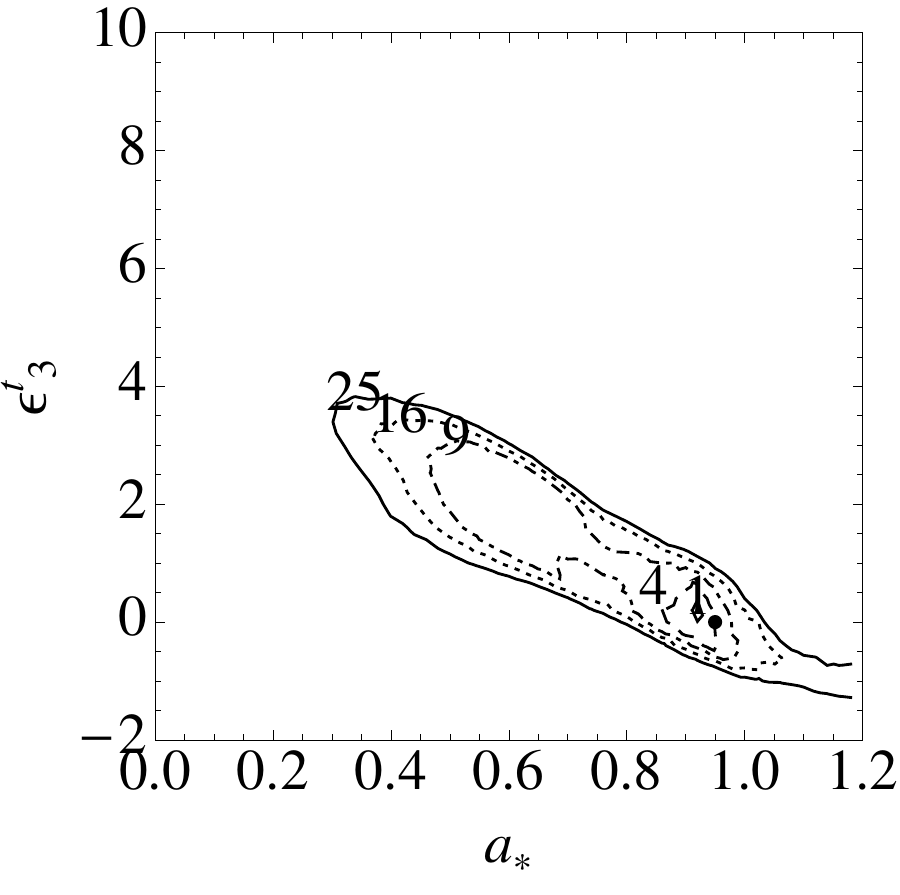} \\
\vspace{0.8cm}
\includegraphics[type=pdf,ext=.pdf,read=.pdf,width=7.0cm]{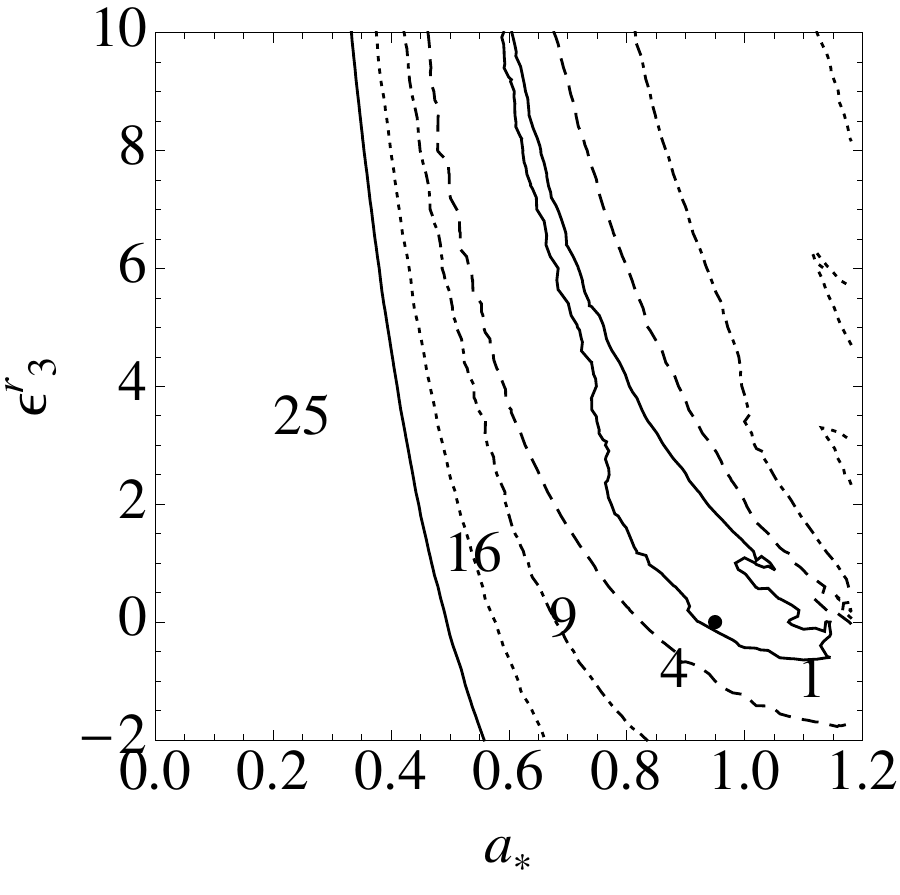}
\hspace{0.5cm}
\includegraphics[type=pdf,ext=.pdf,read=.pdf,width=7.0cm]{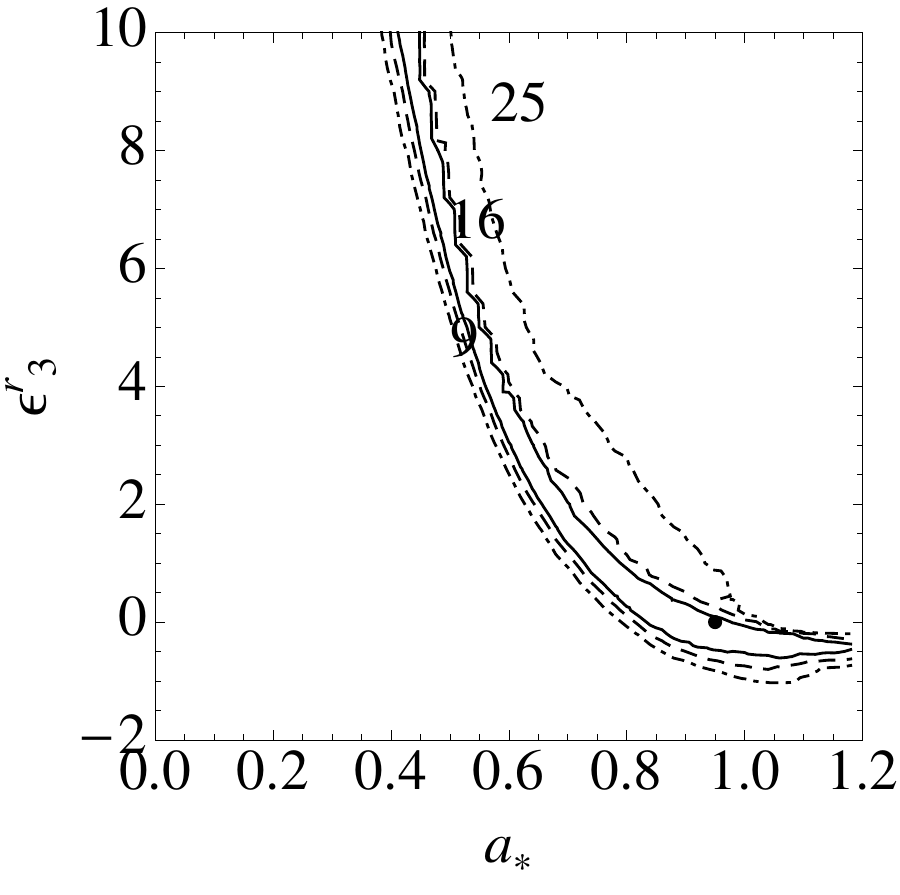}
\end{center}
\caption{Top panels: $\Delta \chi^2$ contours with $N=10^4$ from the comparison of the iron line profile of a Kerr BH with an input spin parameter $a_*' = 0.95$ and an inclination of $i'=20^\circ$ (left panel) and $i' = 70^\circ$ (right panel) versus a set of Cardoso-Pani-Rico BHs with spin parameters $a_*$ and a non-vanishing deformation parameter $\epsilon^t_3$. Bottom panels: as in the top panels for a non-vanishing deformation parameter $\epsilon^r_3$. }
\label{fig2}
\end{figure}

\begin{figure}
\begin{center}
\vspace{0.5cm}
\includegraphics[type=pdf,ext=.pdf,read=.pdf,width=7.0cm]{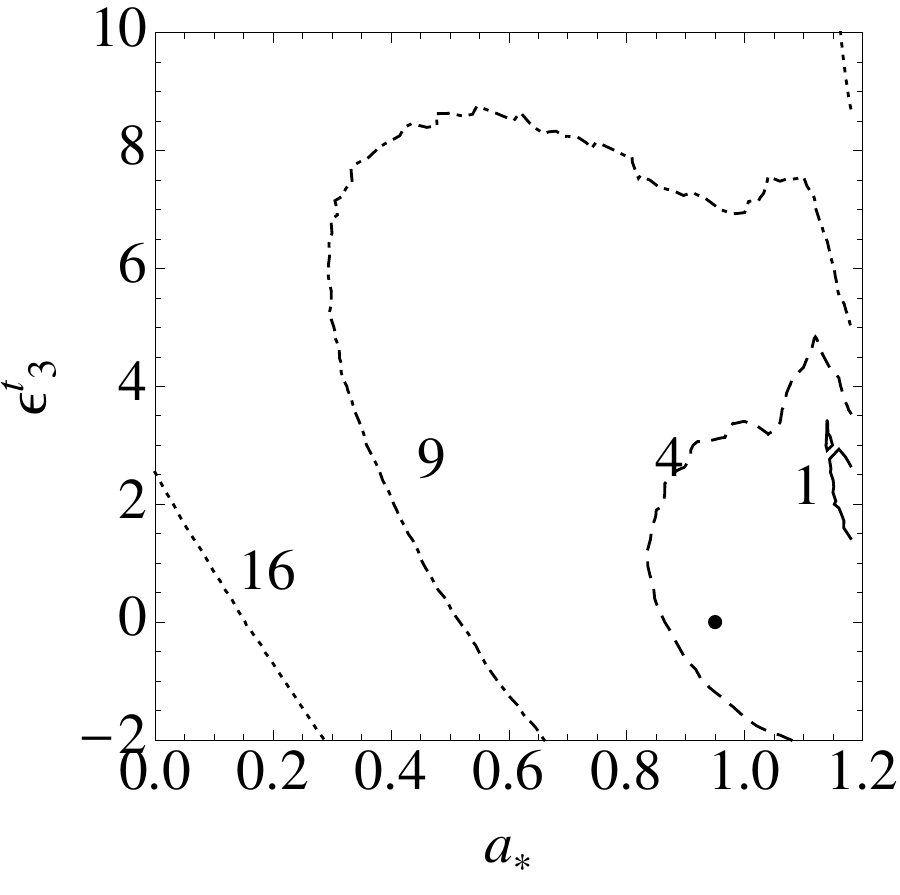}
\hspace{0.5cm}
\includegraphics[type=pdf,ext=.pdf,read=.pdf,width=7.0cm]{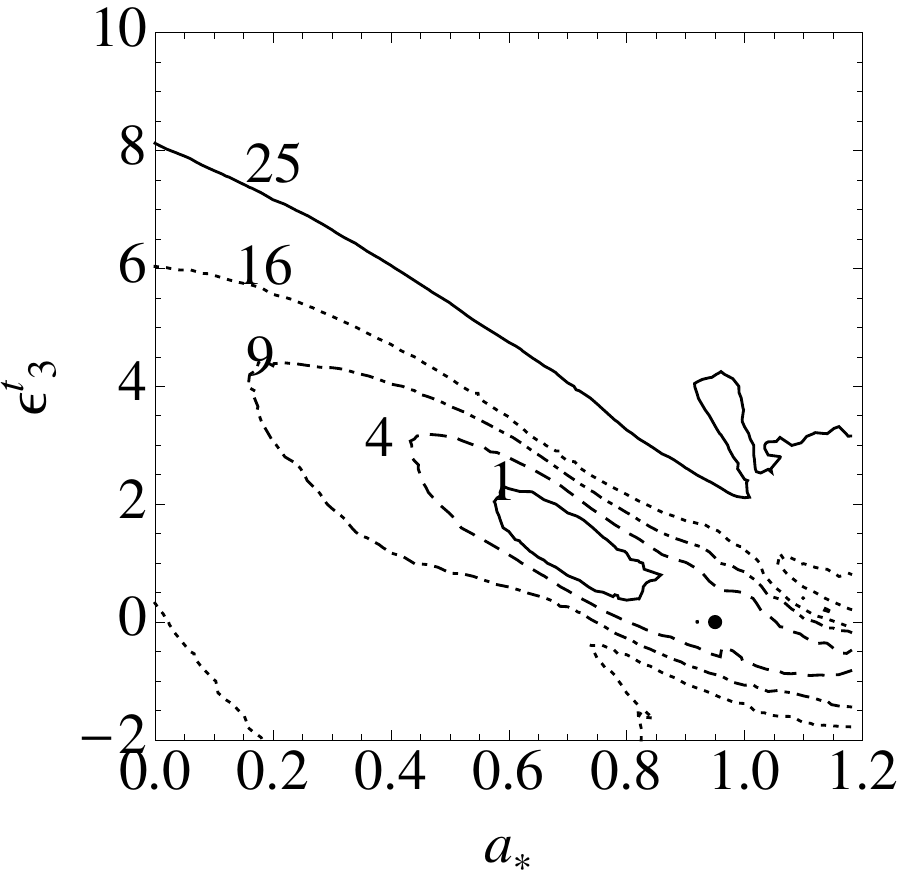} \\
\vspace{0.8cm}
\includegraphics[type=pdf,ext=.pdf,read=.pdf,width=7.0cm]{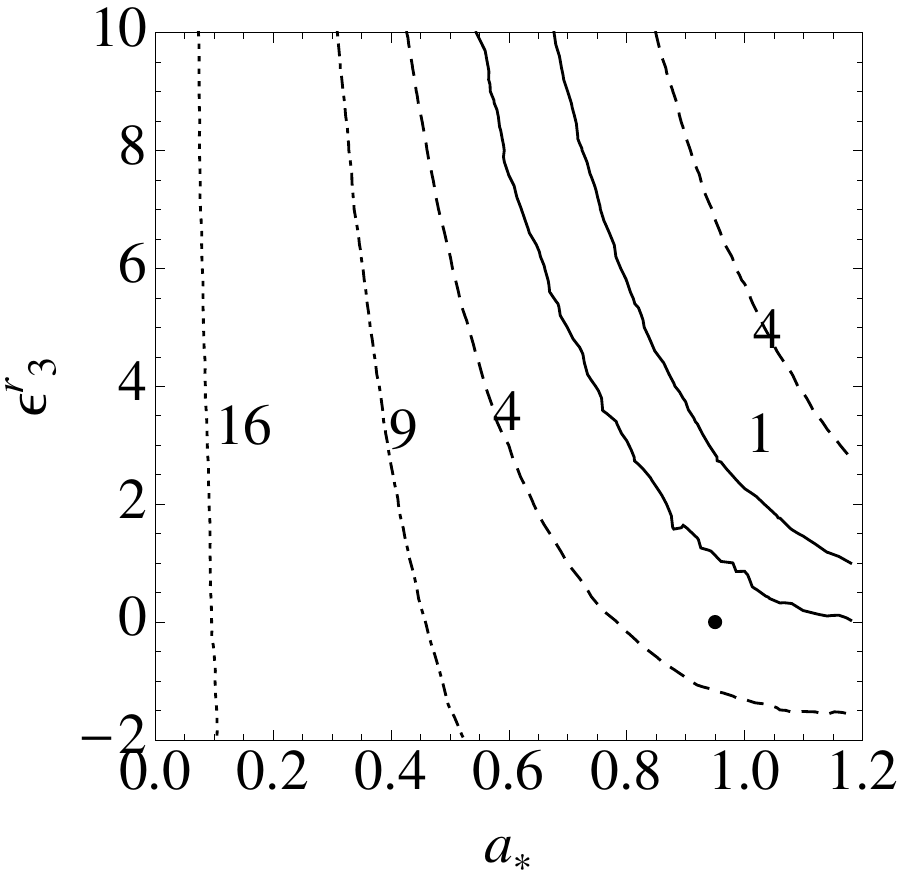}
\hspace{0.5cm}
\includegraphics[type=pdf,ext=.pdf,read=.pdf,width=7.0cm]{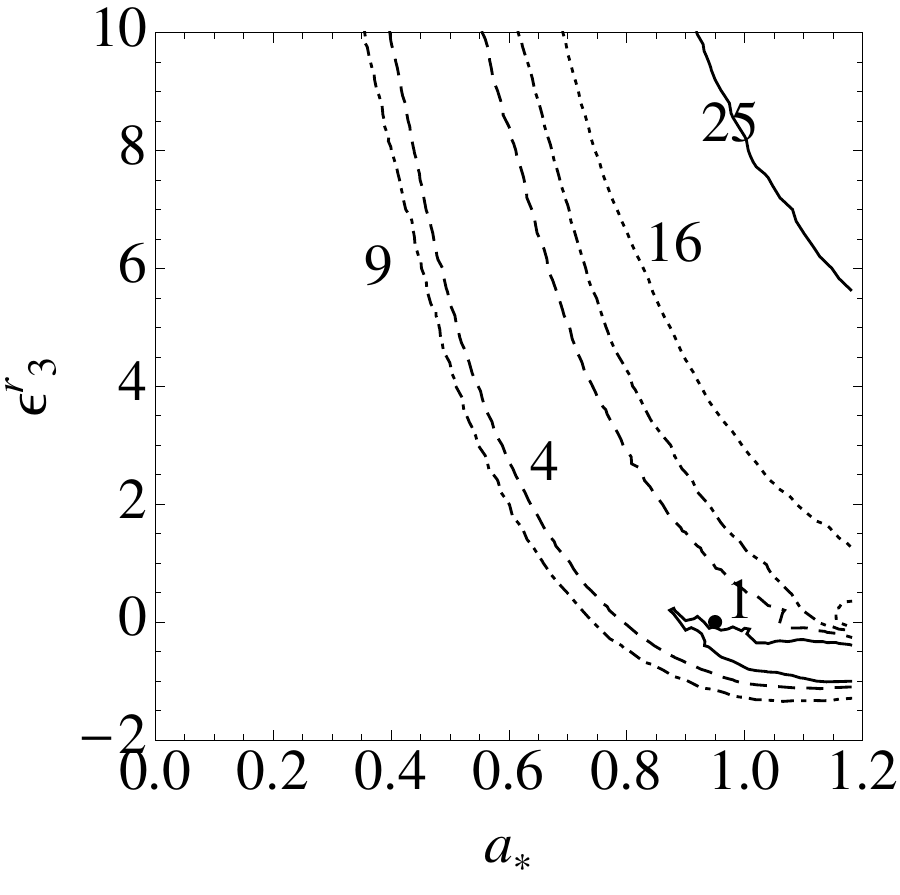}
\end{center}
\caption{As in Fig.~\ref{fig2} for $N=10^3$. }
\label{fig2b}
\end{figure}

\begin{figure}
\begin{center}
\includegraphics[type=pdf,ext=.pdf,read=.pdf,width=7.0cm]{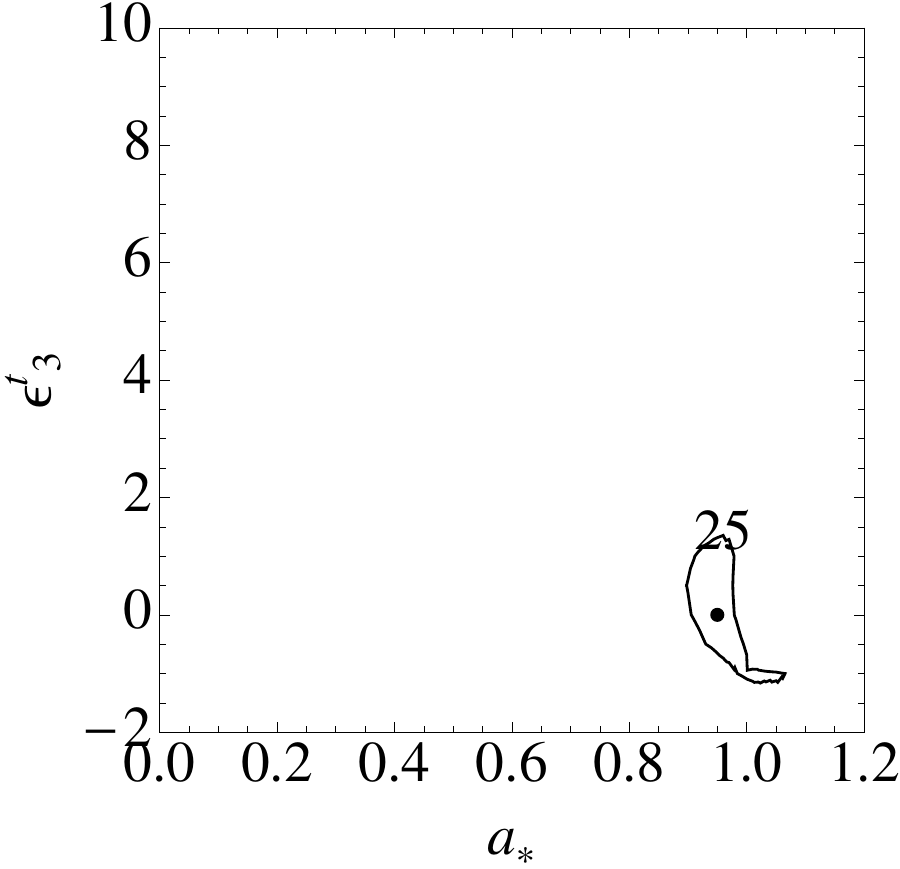}
\hspace{0.5cm}
\includegraphics[type=pdf,ext=.pdf,read=.pdf,width=7.0cm]{0.95-t-70-10000}
\end{center}
\caption{Impact of the power-law continuum. The reference model is a Kerr BH with $a_*'=0.95$ and $i' = 70^\circ$ and the photon number in the iron line is $N = 10^4$. In the left panel, the simulation does not include the power-law component. The right panel, which is the same plot as the top right panel in Fig.~\ref{fig2}, we have both iron line and power-law component. The latter is normalized to include 100 times the number of iron line photons when integrated over the energy range 1--9~keV, and is generated using a photon index $\Gamma'=2$. }
\label{fig2c}
\end{figure}

\begin{figure}
\begin{center}
\vspace{0.5cm}
\includegraphics[type=pdf,ext=.pdf,read=.pdf,width=7.0cm]{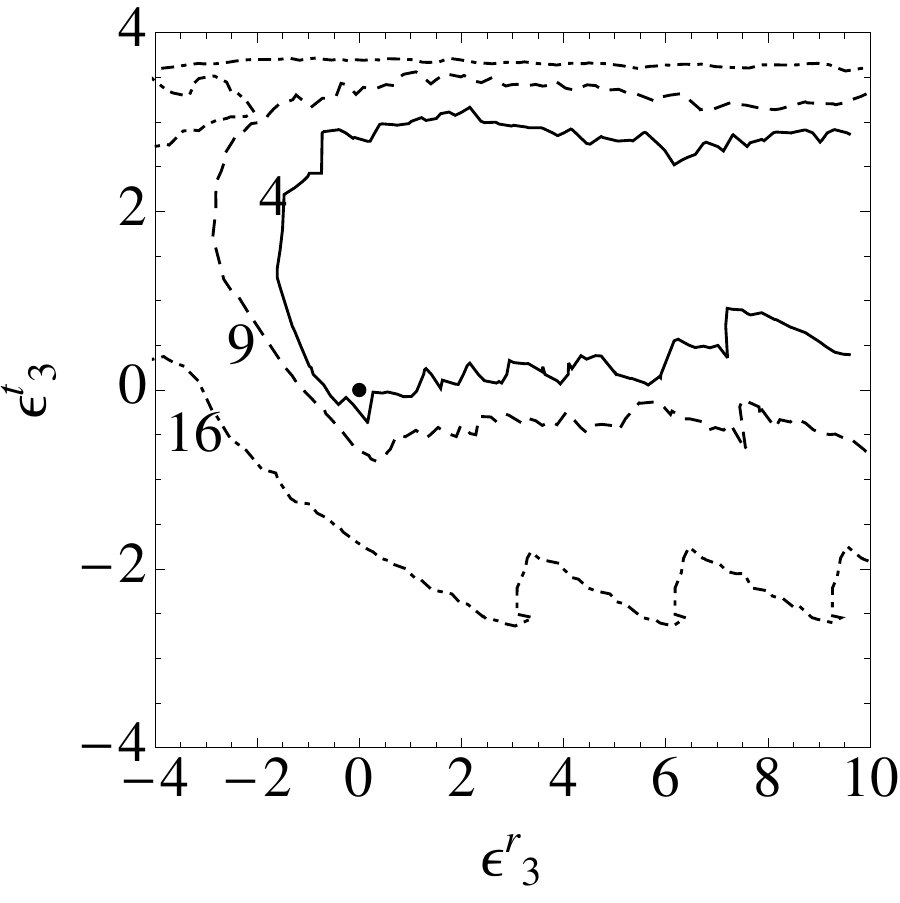}
\end{center}
\caption{$\Delta \chi^2$ contours with $N=10^4$ from the comparison of the iron line profile of a Kerr BH with a spin parameter $a_*' = 0.95$ and an inclination of $i'=70^\circ$ versus a set of Cardoso-Pani-Rico BHs with non-vanishing deformation parameters $\epsilon^t_3$ and $\epsilon^r_3$. See the text for more details.}
\label{fig3}
\end{figure}

Figs.~\ref{fig1}-\ref{fig3} show our results. We begin by considering the two deviation parameters separately, holding one fixed to zero, and allowing the other to vary.  In Fig.~\ref{fig1}, the reference model is a Kerr BH (${\epsilon^t_3}'={\epsilon^r_3}'=0$) with spin parameter $a_*' = 0.5$ and observed from a viewing angle $i'=20^\circ$ (left panels) and $70^\circ$ (right panels). It is compared with a set of models in which the spacetime geometry is described by the Cardoso-Pani-Rico metric with only one non-vanishing deformation parameter, namely $\epsilon^t_3$ in the top panels and $\epsilon^r_3$ in the bottom panels. Here $i$, $K$, and $\Gamma$ are always fit parameters and $\chi^2$ is minimized by varying them. These plots show reference contours of $\Delta\chi^2 = \chi^2 - \chi^2_{\rm min} = 1,$ 4, 9, 16, and 25. It is evident that $\epsilon^r_3$ is more difficult to constrain from the line profile alone.  A similar conclusion is found by \citet{cfm-nk_b}. It is evident that iron lines are more capable of constraining deformation parameters when the inclination of the reference model $i'$ is large, since this maximizes light bending effects.

In Fig.~\ref{fig2}, the reference model is a fast-rotating Kerr BH with spin parameter $a_*'=0.95$. As in Fig.~\ref{fig1}, the left panels are for an inclination angle $i'=20^\circ$ and the right panels are for $i'=70^\circ$, the top panels for a non-vanishing deformation parameter $\epsilon^t_3$ and the bottom panels for a non-vanishing deformation parameter $\epsilon^r_3$. In the right bottom panel, we only show $\Delta\chi^2 = 9$, 16, and 25 for clarity, but we note that $\chi^2$ has an almost degenerate valley extending to to large positive values of $\epsilon^r_3$. $i$ is still free and $\chi^2$ is minimized over $i$. We find that rapidly-rotating BHs make better targets for testing the Kerr metric, enabling the iron line analysis to produce more stringent constraints. Here also, a larger inclination angle is helpful, and again $\epsilon^r_3$ proves much more challenging to measure than $\epsilon^t_3$. Even in the case of a fast-rotating BH with an almost edge-on disk, the iron line cannot really distinguish a Kerr spacetime from a background with a very large $\epsilon^r_3$. A longer exposure time or a larger effective area of the detector, namely a higher number of photons $N$, would surely provide stronger constraints on $\epsilon^t_3$, as $\chi^2$ is approximately proportional to $N$. However, a higher photon count would not be very helpful for $\epsilon^r_3$ because the valley of $\chi^2$ is almost degenerate: the area of the allowed region in Fig.~\ref{fig2} would be smaller, but still large positive values of $\epsilon^r_3$ would be permitted.

Fig.~\ref{fig2b} illustrates how the photon count number in the line $N$, which enters in the Poisson noise, affects the fit constraints. The values of the parameters are the same as in Fig.~\ref{fig2}, but here we have $N = 10^3$ photons in the iron line ($K' = 100$ as before). The constraints are markedly worse. $N = 10^3$ photons corresponds approximately to the line signal available in a present-day high-quality observation of a supermassive black hole, while $N = 10^4$ photons can be expected using the next generation of X-ray satellites. For instance, in the case of a Kerr BH with $a_*'=0.95$ and $i'=70^\circ$, the contour $\Delta\chi^2 = 4$ (9) provides the constraints
\be
& -1 \lesssim \epsilon^t_3 \lesssim 3 & \hspace{0.3cm} (-2 \lesssim \epsilon^t_3 \lesssim 4.5) \;\; {\rm for} \;\; N= 10^3 \, , \nonumber\\
& -0.5 \lesssim \epsilon^t_3 \lesssim 0.5 & \hspace{0.3cm} (-1 \lesssim \epsilon^t_3 \lesssim 3) \;\; {\rm for} \;\; N= 10^4 \, .
\ee
The deformation parameter $\epsilon^r_3$ is unconstrained with both $N=10^3$ and $N = 10^4$. However, for a fixed $\epsilon^r_3$ the uncertainty on the spin parameter is different. The contour $\Delta\chi^2 = 9$ yields a spin uncertainty $\Delta a_* \approx 0.3$ for $N=10^3$, and $\Delta a_* \approx 0.05$ for $N=10^4$.

It is instructive to consider the effect of the power-law continuum on the constraining-power of the iron line signal. This is illustrated in Fig.~\ref{fig2c} for the deformation parameter $\epsilon^t_3$. The reference model is a Kerr BH with $a_*'=0.95$ and $i' = 70^\circ$. The left panel employs the same model as in Fig.~\ref{fig2}, but with just the line, and no power-law component. The right panel is the same plot as the top right panel in Fig.~\ref{fig2}. The continuum's presence adds additional photon noise to the line's signal, which unsurprisingly weakens the resulting constraints, for a given line strength. The contour $\Delta\chi^2 = 25$ would give the constraint
\be
0.9 \lesssim a_* \lesssim 1.1 \, , \quad -1 \lesssim \epsilon^t_3 \lesssim 1.5
\ee
without power-law component. The constraints become  
\be
0.3 \lesssim a_* \lesssim 1.2 \, , \quad -1 \lesssim \epsilon^t_3 \lesssim 4
\ee
if we include the continuum.

Lastly, Fig.~\ref{fig3} shows the case in which both $\epsilon^t_3$ and $\epsilon^r_3$ are allowed to be non-vanishing at the same time. The reference model here is a fast-rotating Kerr BH with spin parameter $a_*'=0.95$ and viewing angle $i'=70^\circ$. According to the previous plots, this is a fruitful source for testing the Kerr metric. Here, spin, inclination, $K$, and $\Gamma$ are optimized in the fit. Strikingly, the measurements of $\epsilon^t_3$ and $\epsilon^r_3$ are essentially uncorrelated.

The constraints obtained in Figs.~\ref{fig1}-\ref{fig3} can be qualitatively understood as follows. When the spin parameter vanishes, $g_{t\phi} = 0$, and the position of the inner edge of the disk is only determined by $\epsilon^t_3$. $\epsilon^r_3$ affects instead the photon propagation from the disk to the distant observer. Since the position of the inner edge of the disk has a strong impact on the iron line profile, $\epsilon^t_3$ is much easier to constrain than $\epsilon^r_3$. For a non-vanishing spin, the picture is more complicated, but the general explanation is still valid and the influence of $\epsilon^r_3$ on the inner edge of the disk is only moderate.

\section{Discussion and conclusions \label{s-c}}

An accurate measurement of the iron K$\alpha$ iron line commonly observed in the X-ray spectrum of both stellar-mass and supermassive BH candidates is potentially a powerful tool for probing the spacetime geometry around these objects and test the Kerr BH paradigm. In this paper, we have extended previous studies by exploring how high quality X-ray data can test the nature of BH candidates for different BH spins, different viewing angles, and two independent types of deviations from Kerr. Assuming the case of a detection of $N=10^4$ photons in the iron line and $N_{\rm C}=10^6$ photons in power-law continuum, we have performed a number of simulations, whose constraints are reported in Figs.~\ref{fig1}-\ref{fig3}. Our results can be summarized as follows:
\begin{enumerate}
\item The best target sources to test the Kerr metric are BH candidates with high spin parameter and observed from a large inclination angle. This is perfectly understandable, as both the situations enhance the magnitude of relativistic effects. 
\item Observations can actually constrain the deformation parameter $\epsilon^t_3$, while constraints on $\epsilon^r_3$ seems to be out of reach with the sole use of iron line data. To measure $\epsilon^r_3$ it is presumably necessary to combine the iron line measurement with a technique only sensitive to the spin parameter, and an example could be the determination of the spin through the jet power~(e.g., as suggested by \citealt{nm-jet, cb2_a}), if such a relationship is confirmed. Since $\epsilon^r_3$ affects the photon propagation, it is also possible that reverberation data, which have a timing information, can help to constrain this parameter.
\item While we have not yet explored the whole parameter space, our results suggest that the measurements of $\epsilon^t_3$ and of $\epsilon^r_3$ are not very correlated, namely it is possible to distinguish the effects of the two classes of deformation on the iron line profile.
\end{enumerate}
In this exploratory work, we have not considered the effects of varying other key parameters related to Fe line modeling (emissivity index $q$, ionization parameter $\zeta$, etc.). Moreover, we have made the simplifying assumption that systematic effects which may otherwise obscure our ability to assess these subtle deviations are under full control. In future work, it will be important to employ a more rigorous treatment of full reflection modeling, to assess systematic effects, and ultimately to perform a full and detailed analysis on the best available data of iron lines produced in strong gravity, in order to constrain any deviation terms directly.


\begin{acknowledgments}
J.J. and C.B. were supported by the NSFC grant No.~11305038, the Shanghai Municipal Education Commission grant No.~14ZZ001, the Thousand Young Talents Program, and Fudan University.
J.F.S. was supported by the NASA Hubble Fellowship grant HST-HF-51315.01.
\end{acknowledgments}

\end{document}